# Knots of Darkness in Atmospheric Turbulence

D. G. Pires[†], D. Tsvetkov[†], N. Chandra, and N. M. Litchinitser[*]

*Department of Electrical and Computer Engineering, Duke University, Durham, North Carolina 27708, USA*

[*]natalia.litchinitser@duke.edu

[†]These authors contributed equally to this work

**Topology, which originated as a mathematical discipline, nowadays advances the understanding of many branches of science and technology from elementary particle physics and cosmology to condensed matter physics. In optics, the topology of light and darkness facilitates new degrees of freedom for sculpting optical beams beyond conventionally used amplitude, phase, and polarization. This fundamentally new, spatial dimension opens new opportunities for several optical applications, ranging from optical manipulation, trapping, data processing, optical sensing and metrology, enhanced imaging, and microscopy, to classical and quantum communications. While topological stability of mathematical knots implying robustness to perturbations suggests their potential as information carriers, the behavior of optical knots in perturbative environments such as atmospheric turbulence is largely unexplored. Here, we experimentally and theoretically investigate the effects of atmospheric turbulence of optical knot stability and demonstrate that the number of crossing (the topological invariant) is preserved in the weak-turbulence regime, but may not be conserved in the stronger turbulence conditions. The turbulent medium is simulated in the laboratory using phase screens, which carry the refractive index changes associated with the Kolmogorov power spectrum, encoded in a spatial light modulator. The optical knots are reconstructed by single-shot measurements of the complex field, and the resilience of the knot topology is analyzed for various realistic turbulence strengths. These studies may give rise to entirely new approaches to the three-dimensional (3D) spatially resolved probing of turbulence.**

Topological concepts such as singularities of the phase, polarization, or energy flow, topological textures, super-oscillations, and spin-orbit interactions once considered mostly in the domains of abstract mathematics or theoretical physics, are now entering the field of classical and quantum optics. Originally, knotted solutions were predicted in the non-paraxial regime by superposing Bessel beams and tracking the phase singularities trajectories along the longitudinal direction [2]. Later, isolated knotted fields were obtained by expanding the Milnor polynomial of a particular mathematical knot unto the Laguerre-Gaussian (LG) basis in both phase and polarization domains [1, 3, 4].

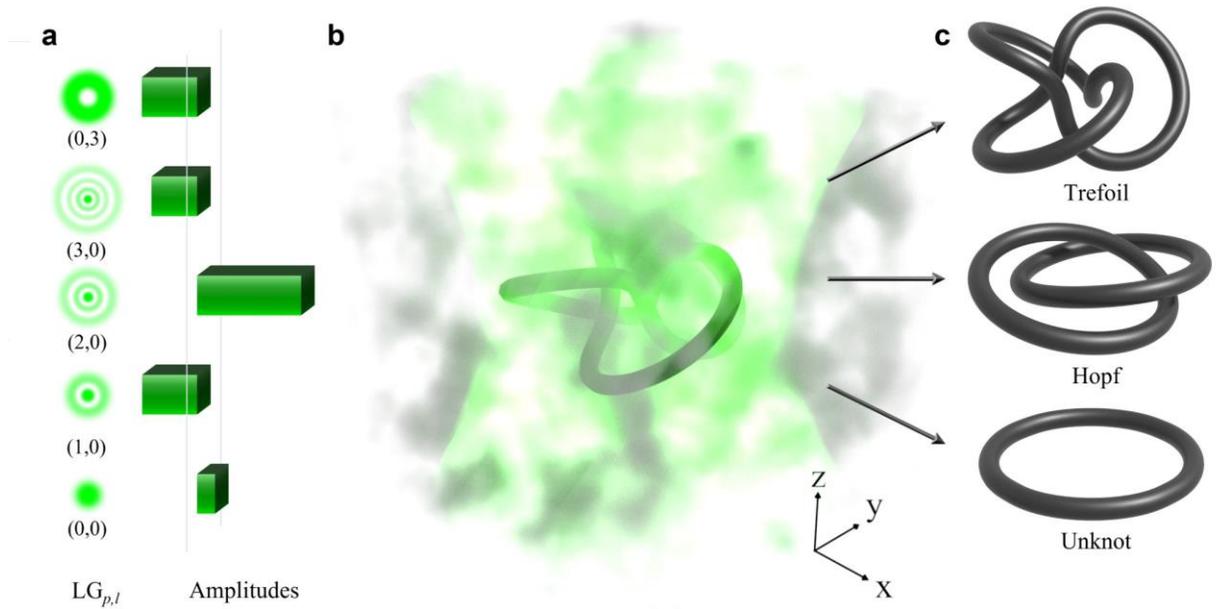

**Fig. 1 Turbulence effect on optical knots.** (a) Superposition of LG beams in a trefoil knot, where $(p, l)$ are the respective indices of the modes. (b) Schematic of the idea presented in this work. In the presence of turbulence, the probability of an originally knotted field (trefoil) gets untied increases. (c) In this case, the trefoil knot may transform itself into a Hopf link or an unknot as it loses crossings due to the refractive index changes inside the medium.

While the mathematical theory of knots and links has been developing over many decades, optical knots behavior in realistic perturbative physical media remains largely unexplored. Understanding the topological stability of knotted solutions is important for their potential applications in the fields of classical and quantum communications [5, 6], microfabrication [7, 8], and quantum computing [9, 10]. In topology, objects (e.g., knots) are equivalent to each other if one can be transformed into the other through continuous deformation without cutting the lines or allowing the lines to pass through themselves [11, 12]. In other words, the singularity lines of the trefoil knot and Hopf link can deform in size and shape, yet maintain the same number of crossings, i.e., topologically invariant. Notably, the number of crossings is just one of the topological invariants defined in pure mathematics, although most commonly referred to in the field of optical or acoustical knots.

Leveraging the topological properties of optical knots for their robustness in turbid and turbulent media is of interest not only from a fundamental science viewpoint but may also yield novel insights in such applications as biomedical imaging, optical manipulation, and probing atmospheric turbulence itself. While the topological stability of optical knots for optical communications in realistic environments has been mentioned in many publications, only recently the first study addressing the robustness of optical knots under specific types of phase aberrations and setup misalignments modeled by the Zernike polynomials has been reported [13]. However, understanding the stability of optical knots in turbulent media remains challenging. Here, we theoretically and experimentally investigate the behavior of optical knots in the presence of atmospheric turbulence and discuss their topological stability in the context of the conventionally used topological invariant and beyond.

**Structured light in atmospheric turbulence.**

Turbulence is a ubiquitous phenomenon that occurs in the atmosphere, the oceans, the wake of a boat, and even in galaxies [14-17]. Atmospheric turbulence leads to beam wandering, scintillation, and most importantly, phase front distortion, which is primarily caused by random temperature variations and convective processes. Although some of these effects can be compensated by using aperture averaging [18], phase distortions are more difficult to counteract. While the exact theoretical modeling of turbulent flows is one of the most challenging problems of modern physics, the Kolmogorov model for turbulent flow accurately describes the turbulence by relating random temperature variations and convective processes that perturb the wavefront of the optical beam to refractive index fluctuations. There are two characteristic parameters used to describe the turbulence: an inner scale, $l_0$, which is typically on the order of millimeters and results in distortions of the wavefront of the beam, and an outer scale, $L_0$, which is on the order of meters and is associated with beam wandering [19]. As for many random processes, theoretical descriptions of turbulence rely on statistical averages for the random variations of the refractive index of the atmosphere. However, experimental results confirm that such statistical descriptions are accurate in most practical cases [20, 21].

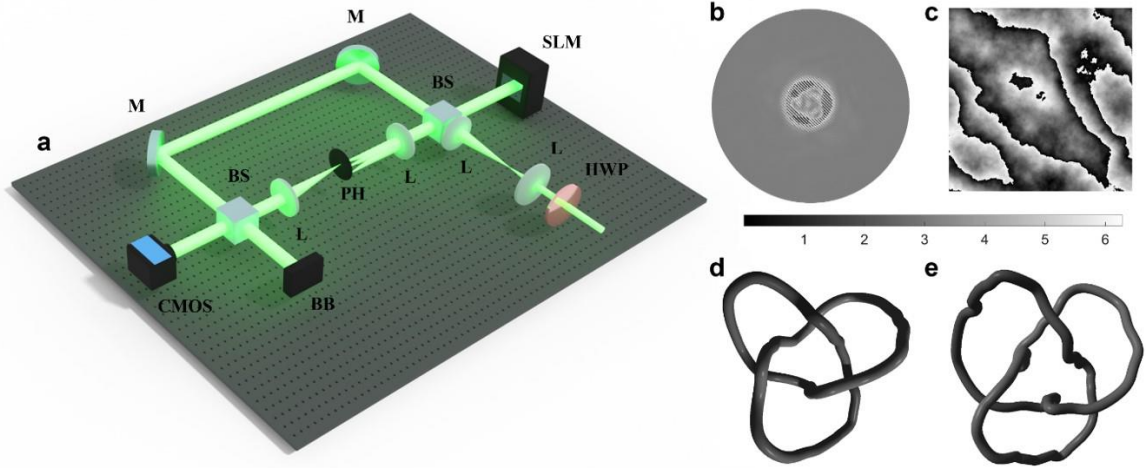

**Figure 2.** (a) In the experiments, a Mach-Zehnder interferometer is used to measure the complex field (amplitude and phase) of the optical beam. Here, HWP = Half-Wave Plate, L = Lens, BS = Beam Splitter, SLM = Spatial Light Modulator, M = Mirror, PH = Pinhole, BB = Beam Blocker, CMOS = Complementary Metal Oxide Semiconductor camera. Examples of a hologram (b) and turbulence phase-screen for SR = 0.9 (c) used in experiments. Theoretical (d) and experimental (e) recovered trefoil knot without turbulence.

In a laboratory environment, a few approaches are commonly used to simulate turbulence. Heaters and fans have been used to emulate randomly changing refractive index distribution in a turbulent chamber by controlling the temperature gradient and air convection [22]. Another technique relies on the random diffusive glass plate that was used as a random-phase screen, where dynamic turbulence can be realized by rotating the plates [23, 24]. However, one of the most versatile techniques is the holographic approach implemented using spatial light modulators (SLM) and digital micro-mirror devices (DMD) with phase screens altered on demand with a refresh rate of up to 32 kHz, mimicking the dynamics of real-world turbulence [25-27]. Therefore, in this study, we used an SLM to encode both knotted field and phase screens, and a single-shot measurement of the complex electric field approach to recover the knots from the phase singularities [28]. Figure 2 shows the experimentally realized Mach-Zehnder interferometer with the SLM in the signal arm to generate the knotted structure together with turbulence phase screens generated from the Kolmogorov power spectrum at a wavelength of 532 nm. The phase screens have been calibrated by recording the on-axis beam intensity with $\langle I(0) \rangle$ and without $I_0(0)$ the turbulence and numerically adjusting the values for the aperture diameter $D$ and the Fried parameter $r_0$ through the Strehl ratio (SR) expression (for more details, see Methods):

$$\text{SR} = \langle I(0)\rangle/I_0(0) \approx \left[1 + (D/r_0)^{5/3}\right]^{-6/5}. \tag{1}$$

**Wandering of optical singularities.**

To understand the parameters that influence the stability of the knot, first, we consider a simpler case of the interactions between a superposition of two $\text{LG}_{p,l}$ beams with variable relative amplitudes and azimuthal and radial indices, $\alpha \text{LG}_{1,0} + \text{LG}_{0,1}$. This case can be studied analytically taking into account both amplitudes and beam widths of the beams, as shown in the Supplementary Materials, Section 1. Figure 3 (a-c) shows the intensity and phase distributions for this field superposition for (a) $\alpha = 1$, (b) $\alpha = 0.7$, and (c) $\alpha = 0.5$, suggesting that as $\alpha$ decreases, the amplitude associated with the radial mode $\text{LG}_{1,0}$ also decreases, leading to a spatial shift of one of the singularities. Next, the propagation of the same fields' superposition was theoretically and experimentally studied in the presence of Kolmogorov turbulence. Figures 3 (d-f) show the experimentally measured probability densities of finding both phase singularities at $(x, y)$ for the same $\alpha$ values and with turbulence strength $\text{SR} = 0.95$ together with the measured phase distributions for each case shown in the insets. These results indicate that the inner singularity experiences less wandering than the outer one as the amplitude of the radial mode decreases. We associate this feature with the surrounding amplitude protecting the optical singularity, and as $\alpha$ decreases, the outer singularity possesses less amplitude surrounding it. It is worth mentioning that the asymmetry on the *y*-direction of the measured positions of the optical singularities (panels d-f) is due to the *z*-location of the camera, where it is slightly off the $z = 0$ position. Figures 3 (g-i) show the trajectory of the singularities along the propagation distance, highlighting that the inner singularity experiences less wandering through the propagation compared to the outer one. These results reveal the link between the relative amplitudes of the LG modes and the relative positions of singularity lines. They can be used as guidelines for the development of an optimization procedure for optical knots and links in turbulence.

**Optical knots.**

Using the analytical expression for the isolated knotted fields [1], we developed a numerical optimization procedure to minimize the perturbations, including both experimental setup misalignments and aberrations, as well as actual turbulence-induced distortions. In the absence of turbulence, Dennis et al. [1] showed that by minimizing of cost function $\sum_{voxels\ in\ volume}[min(I_{sat}, I(\vec{r}))]^{-1}$, where $I_{sat}$ represents the saturated intensity, the accuracy of optical knot measurements can be significantly improved. Following several hundred iterations, this method results in adequately spaced inner singularities and straightened singularity lines along the propagation direction. This process enhances the precision of optical knot measurements and characterization under laboratory conditions. Inspired by this method, we developed an algorithm to maximize the distance between phase singularity points in the 3D space for optical knots in turbulence where the random refractive index perturbations may lead to reshaping and, potentially, disconnections and reconnections of the singularity lines. While our discussion will focus on a particular case of the trefoil knot, the same procedure can be applied to any other knot. The optimization results for the Hopf links are discussed in Supplementary Materials, Section 5.

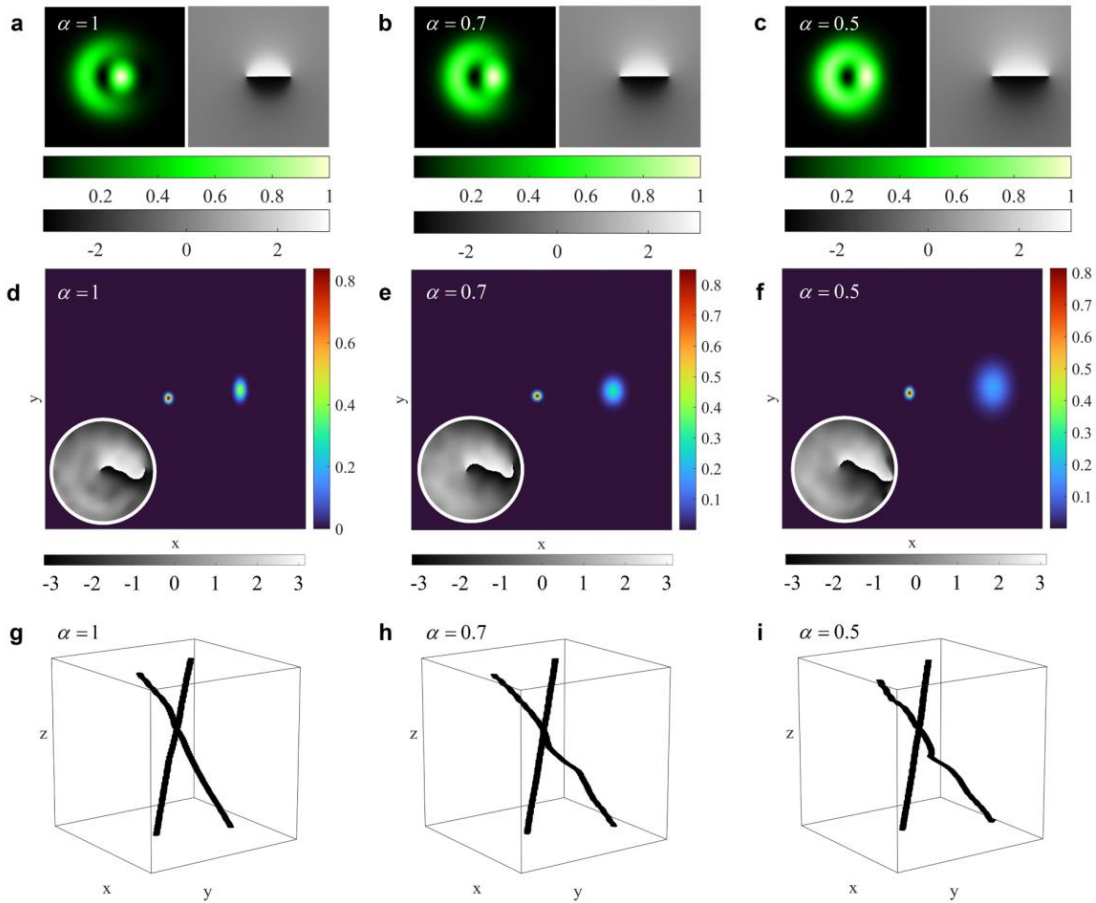

**Fig. 3 Optical singularities wandering in turbulence.** (a-c) Intensity and phase (grey color) distributions for $U = \alpha LG_{1,0} + LG_{0,1}$ with (a) $\alpha = 1$, (b) $\alpha = 0.7$, and (c) $\alpha = 0.5$. (d-f) Experimentally measured probability densities of finding the singularities at $(x, y)$ together with phase distributions (grey color) shown in the insets for the same values of $\alpha$ and turbulence strength SR = 0.95. (g-i) Trajectories of the singularities along the propagation distance.

Starting with a superposition of the LG beams

$$E_{trefoil} = c_{0,0}(w)\text{LG}_{0,0} + c_{1,0}(w)\text{LG}_{1,0} + c_{2,0}(w)\text{LG}_{2,0} + c_{3,0}(w)\text{LG}_{3,0} + c_{0,3}(w)\text{LG}_{0,3}, \quad (3)$$

obtained from the projection of the Milnor polynomial with a fixed Gaussian beam width parameter $w = 1.125$ (see Section 3 of the Supplementary Materials for the details of the optimization process). Subsequently the amplitudes of each individual LG mode in decomposition (3) were swept until the geometry of the knot satisfied the following criteria. First, the topology of the optical knot must stay unchanged, that is, in this case, the knot must remain a trefoil. Second, extra singularities that may appear due to the Gaussian envelopes of LG beams, as discussed below, must be located at a distance no less than the preset minimum distance from the isolated knot along the propagation direction. We found that setting this minimum distance to be 25% of the total knot length along the direction of propagation was enough to prevent the merging of the knot with these extra singularities due to turbulence. Third, the values of amplitudes of the LG modes at a particular iteration step are kept if the value of the cost function $\left[min(|\vec{r}_i^k - \vec{r}_j^k|)\right]^{-1}$ decreased as compared to the previous step, otherwise, the procedure of varying these amplitudes continues. In the expression of the cost function $k$ corresponds to different transverse $(x, y)$ cross-sections along the knot, while $i$ and $j$ are integers corresponding to all singularities in a particular $k$-plane assuming $i \neq j$, $\vec{r}_i^k$ and $\vec{r}_j^k$ are the vectors corresponding to the points where the singularity lines constituting the knot cross a particular $k$-plane (see Fig. 4). Note that in some planes, extra singularity crossing points may appear due to the bending of the singularity line (*e.g.*, on either end of the knot where a single line crosses the transverse plane twice as shown in plane $k_2$ in Fig. 4 (d), (f)). In this case, not all pairs of points should be considered when calculating the cost function. If these points do not lead to a change of topology, we skip the distance calculation between them. Here, by a particular topology, we understand the number of crossings possessed by a certain recovered structure. The sweeping of the LG modes amplitudes required to build the optimized trefoil knot was performed within the range of 3% of the current maximum coefficient value and took a few hundred iterations to be completed. Figure 4(d) presents the structure of the optimized trefoil knot, juxtaposed with the structure obtained using the standard mathematical expression (3), as shown in Fig. 4(a).

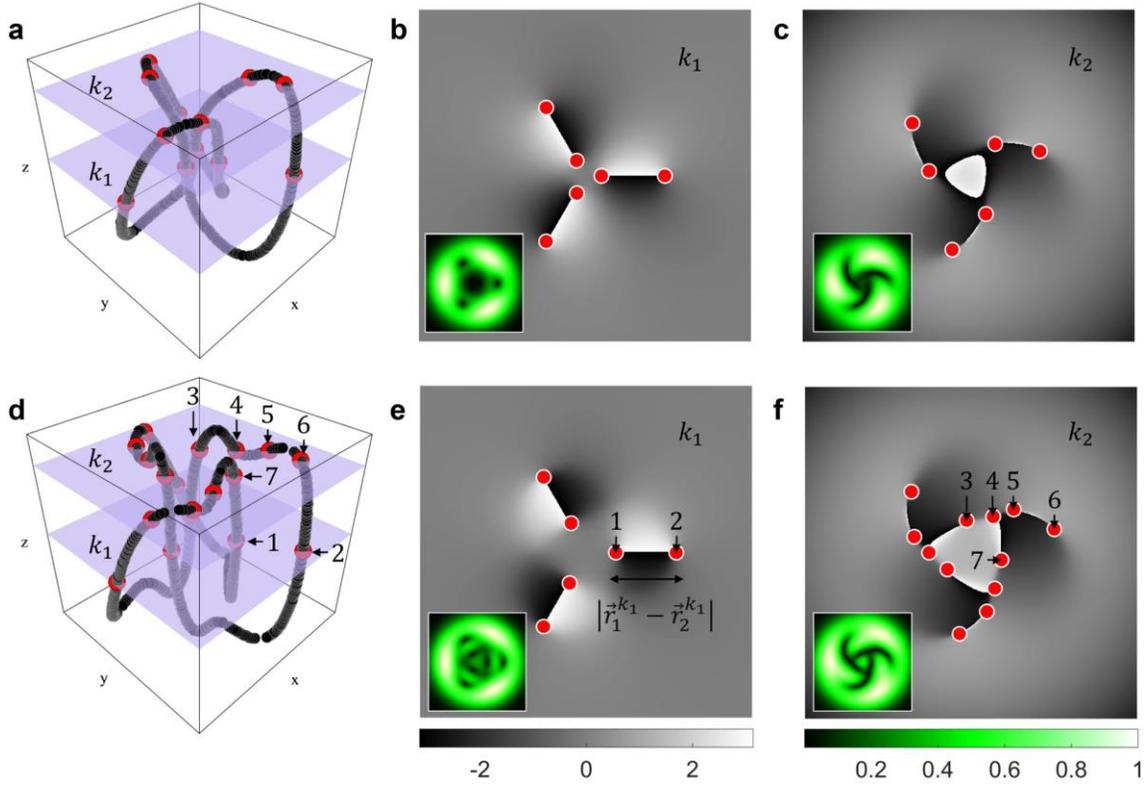

**Fig. 4 Optimization scheme of knotted fields.** (a-c) Unoptimized and (d-f) optimized trefoil knots sliced in two different cross-sections $k_1$ and $k_2$. Phase distribution at $k_1$ (b,e) and $k_2$ (c,f) planes, where an example of the distance between two singularities $|\vec{r}_1^{k} - \vec{r}_2^{k}|$ is illustrated in panel (e) and their respective amplitude distribution is shown in the inset. Red dots indicate the points where the singularity lines cross a particular $k$-plane at each plane. In panel (f) we highlight that, even though the singularities 3-6 are close to each other, their connection does not change the topology of the knot. At this plane, the algorithm optimizes the distance between 7 and the singularities 3-6. The coefficients for the unoptimized [optimized] optical trefoil knot for each LG modes $(p, l)$ are: (0,0) 1.71 [1.29]; (1,0) -5.66 [-3.95]; (2,0) 6.38 [7.49]; (3,0) -2.3 [-3.28]; (0,3) -4.36 [-3.98], where all the coefficients are normalized so the sums of the squares are equal to 100.

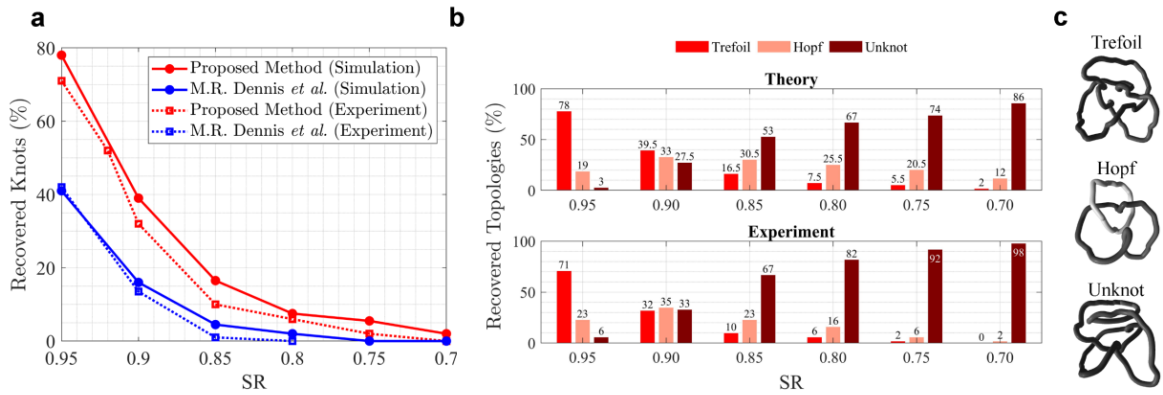

**Fig. 5 Stability of recovered knots.** (a) The percentage of numerically (dots) and experimentally (squares) recovered trefoil knots obtained using our proposed coefficients' optimization procedure (red) as a function of the strength of the turbulence (SR) compared to the previously reported saturated intensity-based optimization technique [1] (blue). (b) Histograms showing the recovered trefoil knots, Hopf links, and unknots for various turbulence strengths using the optimization proposed in the manuscript. (c) Examples of experimentally recovered trefoil knot, Hopf link, and an unknot, respectively.

**Stability of optical knots.**

The refractive index changes from turbulence modify the coefficients in the LG superposition, thus altering the topology of the singularity trajectory lines. As the turbulence becomes stronger, the probability of the original knot losing crossings increases. For instance, a trefoil knot tends to change its topology to either a Hopf link or unknot, as shown in Fig. 5. After generating approximately 200-300 different phase screens for each value of the SR parameter, we reconstruct the optical knot in the far field and statistically calculate the percentage of recovered knots for both theoretical and experimental scenarios. Here, we include the turbulence phase screen to the inverse Fourier transform of the optical field and then it is propagated to the far-field, where the disturbed field is measured by a camera positioned at the image plane of the SLM. After measuring the complex field at a single plane, the three-dimensional field is recovered by using the angular spectrum method to numerically propagate it in both forward and backward propagation directions [28]. Subsequently, the phase singularities are identified in each propagation plane, and their connections are established. This method improves precision in determining the positions of singularities, coupled with rapid, automated phase measurements. The topological structure of these singularities is then classified, as detailed in Section 4 of the Supplementary Materials. Additional examples of the recovered topologies are shown in Section 6 of the Supplementary Materials.

Figure 5(a) shows the percentage of numerically (dots) and experimentally (squares) recovered trefoil knots as a function of the strength of the turbulence obtained using our proposed coefficients' optimization procedure compared to the performance of the previously published technique [1]. Figure 5(b) shows the percentage of recovered trefoil knots, Hopf links, and unknots for each turbulence strength. To categorize the final structure correctly, we incorporate a Python package [29] in our analysis responsible for calculating the Alexander polynomial for each knot. Since this is a knot invariant, we can identify whether the final knot is a trefoil, Hopf link, or unknot. These results quantify the stability of the knots as a function of the strength of the turbulence and can be understood in the context of the results shown in Fig. 2. The developed optimization procedure results in modal amplitudes (relative weights) that maximize the intensity of light surrounding the singularities composing each optical knot leading to a higher percentage of recovered knots in atmospheric turbulence.

**Discussion**

So far we experimentally and theoretically demonstrated that optical knots optical knot stability characterized by the number of crossing is preserved in the weak-turbulence regime, but may not be conserved in the stronger turbulence conditions suggesting that mathematical topological stability of knots may not guarantee the stability of their optical counterparts. Furthermore, we show that the stability of optical knots can be improved by optimizing their constitutive parameters such as amplitudes and beam waists. This conclusion is of paramount importance for their potential applications in the fields of classical and quantum communications [5, 6], microfabrication [7, 8], and quantum computing [9, 10]. It is worth discussing a statement previously mentioned while studying the wandering of optical singularities. We have addressed those singularities surrounded by higher amplitude contrast experience less wandering when compared to lower amplitude contrast. Both optimization methods considered in this work lead to higher amplitude contrast around the multiple singularities disposed throughout the optical field. Originally, the optimization developed by Dennis *et.al.* [1] meant to facilitate the experimental measurements and we observe that, in the presence of atmospheric turbulence, increasing the amplitude contrast leads to less singularity wandering, and thus to higher stability for three-dimensional singularity lines. We stress that further studies must be conducted in this direction and might lead to the improvement of optical communication protocols using OAM-carrying beams by designing its complex field aiming for higher contrast surrounding the optical singularities.

In conclusion, our results indicate that the topological structure of optical knots is relatively stable in the weak-turbulence regime. It is worth mentioning that SR values considered in Fig. 5 corroborate with the expected, realistic measured values of turbulence strengths in free-space without the presence of rain [30]. However, post- or pre-correction algorithms may need to be added to maintain the shape of the knot in strongly perturbed environments such as strong-turbulence regime [31], calling into question the topological stability of optical knots in terms of the conventional topological invariant related to the number of singularity crossings. Nevertheless, the topological properties of optical knots become evident after examining the reconnection events due to turbulence. It is noteworthy that the knots and links considered in this study are formed after the constituent LG beams are transmitted in turbulence. Therefore, depending on the strength of the turbulence, the LG beams interfere in 3D space to form a deformed 3D interference pattern that we then classify as a trefoil knot, Hopf link, or an unknot based on the number of crossings. Even in the case when the number of crossings changes, overall 3D patterns closely resemble the targeted optical knot and a link suggesting that the classification of the knots formed in turbulence relying on counting the number of crossings may not be an ideal approach in such complex medium and, perhaps, a different invariant/criterion needs to be defined to classify the knots in turbulent media. It is expected that using such a 3D shape-based criterion will result in a yet larger percentage of original knots or links being recovered after turbulence. The limitations of the proposed optimization and classification methods are further discussed in Section 7 of the Supplementary Materials. Once the new and more efficient criterion is established, knotted fields can be further considered to carry information for communication systems, probing the turbulence itself, as well as for metrology and sensing applications. The concept behind this study might be also important to other physical systems, such as Bose-Einstein condensates, fluids, and quantum optics, to cite a few.

**Acknowledgments**

This work was supported by Multidisciplinary University Research Initiative (Grant Number: N00014-20-2558) and Army Research Office (W911NF2310057).

**Methods**

**Encoding holograms and phase recovery.** Optical knots that are the null solutions of the Helmholtz equation in three-dimensional free space have only recently been demonstrated in optical experiments through overlapping multiple LG beams with carefully chosen relative amplitudes in 3D space such that their 3D interference pattern consists of closed loops of complete darkness that can be described as

$$E_{knot}(r) = \sum_{p,l} c_{p,l} \text{LG}_{p,l}(r, w_0), \qquad (2)$$

where the sum is taken over all modes where $c_{p,l} \neq 0$, and each term satisfies the linear paraxial Helmholtz equation. Here, $c_{p,l}$ are the coefficients of the respective $\text{LG}_{p,l}$, where $p$ and $l$ are the radial and azimuthal indices, respectively, $w_0$ is the LG beam width parameter that is assumed to be the same for all $\text{LG}_{p,l}$ modes (see the Supplementary Materials (S2)). After encoding the holograms into the SLM using an approximated inverse *sinc*-function method [32], the phase distribution $\phi$ and amplitude $A$ are measured by numerically inverting the expression $\tan(\phi) = (I_4 - I_2)/(I_1 - I_3)$ and $A = \sqrt{(I_4 - I_2)^2 + (I_1 - I_3)^2}$. Here, $I_i$ ($i = 1,2,3,4$) is phase-shifted interferograms equally spaced by $\pi/2$ [33].

**Simulating turbulence using SLMs.** The approach described below translates the changes in the refractive index of the atmosphere, which is caused by fluctuations in the temperature, into phase-only perturbations. After propagation, these phase-only perturbations disturb both the amplitude and phase of the optical field. This thin-phase screen approximation is particularly interesting for condensing the turbulent media into a single-phase screen [34].

To model the random phase screen $\Phi$ associated to the atmospheric turbulence, we define the following phase structure function

$$\Psi(r) = \langle |\Phi(r' + r) - \Phi(r')|^2 \rangle, \qquad (3)$$

where $\langle \cdot \rangle$ refers to the ensemble average, and we relate the Weiner spectrum $\Theta$ to the phase structure function as

$$\Psi(r) = 2 \int \Theta(k)[1 - \cos(2\pi k \cdot r)]d^2k. \qquad (4)$$

One can write the above expressions for Kolmogorov spectrum as [35]

$$\Psi(r) = 6.88 \left(\frac{|r|}{r_0}\right)^{5/3}, \tag{5}$$

and,

$$\Theta(k) = \frac{0.023}{r_0^{5/3}} |k|^{-11/3}, \tag{6}$$

where $r_0$ is the Fried parameter. In other words, the Wiener spectrum $\Theta$ is the covariance function of the random power spectrum $P(k)$, which is defined as

$$P(k) = \mathcal{F}\{\Theta\} = \int \Theta \exp[-i2\pi k \cdot r] \, d^2r, \tag{7}$$

with $\mathcal{F}$ denoting the Fourier transform operator. Now, we calculate the random power spectrum $P(k)$ by sampling a random distribution with zero mean and variance equal to $\Theta(k)$ [36]. This leads to a Kolmogorov Wiener spectrum with $N \times N$ entries denoted by

$$\Theta(i,j) = 0.023 \left(\frac{2D}{r_0}\right)^{5/3} (i^2 + j^2)^{-11/3}, \tag{8}$$

where $D$ is the aperture size which $\Phi$ is being calculated and $i, j$ are the indices of each entry of the Kolmogorov spectrum in a square grid. Multiplying $\sqrt{\Theta}$ by a random matrix $M$, composed of complex values with zero mean and variance 1, leads to the power spectrum $P(k)$ which, after taking the inverse Fourier transform, results in the desired phase screen $\Phi$.

Now, as the smallest generated frequency is $1/D$, this does not accurately describe the phase screen dependency on $|k|^{-11/3}$. These subharmonic contributions are considered by evaluating the power spectrum at those frequencies with periods greater than $D$ (or frequencies smaller than $1/D$). Finally, the phase screen is obtained by taking the real part as

$$\Phi = \Re\left\{\mathcal{F}^{-1}\{M\sqrt{\Theta}\}\right\}, \tag{9}$$

now including the subharmonic contributions.

For the Kolmogorov spectrum, the generated phase screens' turbulence strength follows Eq. (1) in the main text. The Strehl ratio (SR) and the dimensionless parameter $D/r_0$ are directly linked, where whether one characterizes the system using the former or the latter makes no difference. For instance, $D/r_0 = 0.238$ leads to $SR = 0.9$. One can also experimentally retrieve the $SR$ of a turbulent system by measuring the ratio of the average on-axis intensity with ($\langle I(\mathbf{0})\rangle$) and without

($I_0(\mathbf{0})$) turbulence. Stronger turbulence will induce higher beam wandering, leading to lower on-axis intensity of the perturbed optical beam, while the opposite occurs for weaker turbulence. This method is also used to experimentally calibrate the above numerically generated phase screens by adjusting the parameter $D$.

**Contributions**

N.M.L., D.G.P., D.T., and N.C. generated the idea and performed theoretical and numerical analysis. D.G.P. built the experimental setup and performed the experiments. D.T. devised the optimization algorithm for optical knot stability and analyzed the topological structures. N.C. developed the analytical model describing the singularity positions for the case of the combination of two LG beams. N.M.L. supervised the project. All authors contributed to the discussion and interpretation of the results and writing of the manuscript.

**Competing interests**

The authors declare no competing interests.

# Supplementary Materials

# Knots of Darkness in Atmospheric Turbulence

D. G. Pires[†], D. Tsvetkov[†], N. Chandra, and N. M. Litchinitser[*]

*Department of Electrical and Computer Engineering, Duke University, Durham, North Carolina 27708, USA*

## S1. Superposition Laguerre-Gaussian modes carrying azimuthal and radial indices

In this section, we obtain analytical expressions for the location of optical singularities in a field superposition in the absence of turbulence. Let's consider a superposition of the Laguerre-Gaussian (LG) modes $LG_{p,l}$. The paraxial wave equation describes a wave that propagates in a highly directional manner along an axial direction. In this study, the beams propagate along the z-axis. The electric field for a linearly polarized wave is given by $\boldsymbol{E}_{p,l} = \hat{\boldsymbol{x}} U_{p,l}^{\text{LG}}(r, \phi, z) e^{-ikz}$. The spatial structure at any axial position depends on the following four quantities: the radius of curvature $R_{p,l}(z) = z + \frac{k^2 w_{0_{p,l}}^2}{2z}$, beam waist $w_{p,l}(z) = w_{0_{p,l}}\sqrt{1 + \frac{4z^2}{k^2 w_{0_{p,l}}^4}}$, Gouy phase $\theta_{G_{p,l}}(z) = \tan^{-1}\left[\frac{2z}{kw_{0_{p,l}}^2}\right]$, the amplitude $U_{p,l}$.

Laguerre-Gaussian modes with zero angular order ($l = 0$) are given by

$$\mathrm{LG}_{p,0} = U_{p,0}\sqrt{\frac{2}{\pi}\frac{1}{w_{p,0}(z)}} L_p^0\left[\frac{2r^2}{w_{p,0}^2(z)}\right] e^{\left(-\frac{r^2}{w_{p,0}^2(z)} + i\left[\frac{kr^2}{2R_{p,0}(z)} - (2p+1)\theta_{G_{p,0}}\right]\right)}, \tag{S1}$$

where the associated Laguerre polynomial for the first radial and zero angular mode $L_1^0(x) = -x + 1$. The associated Laguerre polynomial corresponding to the zero radial order is $L_0^{|l|}(x) = 1$. Laguerre-Gaussian modes with zero radial order ($p = 0$) are given by

$$\mathrm{LG}_{0,l} = U_{0,l}\sqrt{\frac{2}{\pi|l|!}\frac{1}{w_{0,l}(z)}}\left[\frac{\sqrt{2}r}{w_{0,l}(z)}\right]^{|l|} e^{\left(-\frac{r^2}{w_{0,l}^2(z)} + i\left[l\phi + \frac{kr^2}{2R_{0,l}(z)} - (|l|+1)\theta_{G_{0,l}}\right]\right)} \tag{S2}$$

The superposition of both families of modes in the $z = 0$ plane becomes

$$\mathrm{LG}_{p,0}(z=0) + \mathrm{LG}_{0,l}(z=0) = U_{p,0}\sqrt{\frac{2}{\pi}\frac{1}{w_{p,0}(0)}} L_p^0\left[\frac{2r^2}{w_{p,0}^2(0)}\right] e^{-\frac{r^2}{w_{p,0}^2(0)}}$$
$$+ U_{0,l}\sqrt{\frac{2}{\pi|l|!}\frac{1}{w_{0,l}(0)}}\left[\frac{\sqrt{2}r}{w_{0,l}(0)}\right]^{|l|} e^{-\frac{r^2}{w_{p,0}^2(0)}} e^{il\phi} \tag{S3}$$

The positions of the singularities in the $z = 0$ can be found from the equality $\mathrm{LG}_{p,0}(z=0) + \mathrm{LG}_{0,l}(z=0) = 0$, that leads to the following expression:

$$\frac{1}{\sqrt{|l|!}}\left(\frac{\sqrt{2}r}{w_{0,l}(0)}\right)^{|l|} e^{i(l\phi \pm \pi)} = \alpha\frac{w_{0,l}(0)}{w_{p,0}(0)} L_p^{|0|}\left[\frac{2r^2}{w_{p,0}^2(0)}\right] e^{-r^2\left[\frac{1}{w_{p,0}^2(0)} - \frac{1}{w_{0,l}^2(0)}\right]}, \tag{S1}$$

where $\alpha = \frac{U_{p,0}}{U_{0,l}}$. We have separated the contributions from radial and azimuthal mode numbers on either side of the equations.

In the case of $\mathrm{LG}_{01}$ and $\mathrm{LG}_{10}$ we have:

$$\left[\frac{\sqrt{2}r}{w_{0,1}(0)}\right]e^{i(\phi\pm\pi)} = \alpha\frac{w_{0,1}(0)}{w_{1,0}(0)}\left[1 - \frac{2r^2}{w_{1,0}^2(0)}\right]e^{-r^2\left[\frac{1}{w_{1,0}^2(0)} - \frac{1}{w_{0,1}^2(0)}\right]}. \qquad (S5)$$

Assuming both optical modes have the same amplitudes ($\alpha = 0$), beam waists ($w_{0_{1,0}} = w_{0_{0,1}} = w_0$), and the angular coordinate of the phase singularities at $\phi = \pm\pi$, the singularities will occur at $r = 1.1441w_0$ and $r = -0.437w_0$.

Now, considering $\alpha \neq 1$, i.e. $\alpha\text{LG}_{1,0} + \text{LG}_{0,1}$, we find that the singularities are located at $r_1 = w_0(1 - \sqrt{4\alpha^2 + 1})/2\sqrt{2}\alpha$ and $r_2 = w_0(1 + \sqrt{4\alpha^2 + 1})/2\sqrt{2}\alpha$. Using the L'Hôpital's rule we find that as $\alpha \to 0$, the beams' superposition converges to an $\text{LG}_{01}$ mode and the position of the singularities tend to $r_1 \to 0$ and $r_2 \to \infty$. On the other hand, as $\alpha \to \infty$ the radial mode $\text{LG}_{1,0}$ dominates and the position of the singularities converge to $r_1 = -w_0/\sqrt{2}$ and $r_2 = w_0/\sqrt{2}$. These singularities located along the x-axis are related to the ring-shaped singularities.

## S2. Trefoil Knot

Following the procedure outlined in Supplementary Ref. [1], the Milnor polynomial for a trefoil knot can be written as

$$f_{trefoil}(R,\phi,z) = \{(R^4 - 1)(R^2 - 1) - 8R^3 e^{3i\phi}\} + 4iz(R^4 - 1) \qquad (S6)$$
$$+ z^2(3R^4 - 6R^2 - 5) + 8iz^3R^2 + z^4(3R^2 - 5) + 4iz^5 + z^6.$$

To derive the expression for the optical trefoil knot used in both the experimental studies and numerical simulations, we first set $z = 0$ in (S6) (see Fig. S1 (a)). Next, we multiply this expression by the Gaussian envelope given by $\text{LG}_{0,0}(R,\phi,0) = e^{-R^2/2w^2}/(w\sqrt{\pi})$ and obtain

$$F_{trefoil}(R,\phi,0) = \frac{(R^4 - 1)(R^2 - 1) - 8R^3 e^{3i\phi}}{w\sqrt{\pi}} e^{-\frac{R^2}{2w^2}}. \qquad (S7)$$

Figures S1(b-f) show that as the envelope width parameter $w$ increases, the positions of singularities stay the same, but the intensity distributions around the singularities change. By projecting Eq. (S7) onto the LG basis, we can find the coefficients $c_{p,l}(w)$ in Eq. (2) of the main text.

$$\mathrm{LG}_{p,l}(R,\phi,z) = \sqrt{\frac{p!}{\pi(|l|+p)!}} \frac{R^{|l|}e^{il\phi}}{w_0^{|l|+1}} \frac{(1-iz/z_R)^p}{(1+iz/z_R)^{p+|l|+1}}$$
$$e^{-\frac{R^2}{2w_0^2(1+iz/z_R)}} L_p^{|l|}\left(\frac{R^2}{w_0^2(1+z^2/z_R^2)}\right). \quad (S8)$$

Here, $z_R = kw_0^2$ is the Rayleigh length and $L_p^{|l|}$ is the associated Laguerre polynomial. Note that in general, different $\mathrm{LG}_{p,l}$ modes can have different waists $w_{0_{p,l}}$, and in this case, $w_{0_{p,l}}$ can potentially be used as additional parameters to control the relative positions of the singularities. Even though $w_0$ is not necessarily equal to $w$ in Eq. (S7), for simplicity here we set $w_0 = w$. Consequently, the coefficients $c_{p,l}(w)$ corresponding to a particular $\mathrm{LG}_{p,l}$ mode in the superposition, can be determined from

$$\overbrace{1 - w^2 - 2w^4 + 6w^6}^{(p,l)=(0,0)} \bigg| \overbrace{w^2(1 + 4w^2 - 18w^4)}^{(1,0)} \bigg| \overbrace{-2w^4(1 - 9w^2)}^{(2,0)} \bigg| \overbrace{-6w^6}^{(3,0)} \bigg| \overbrace{-8\sqrt{6}w^3}^{(0,3)}, \quad (S9)$$

where $(p,l)$ are the LG mode indices. It is important to note that $w$ the parameter is not identical to the final beam waist of the LG beams in Eq. (2). While the parameter $w$ dictates the intensity distribution around the singularity lines, the LG beam waist $w_0$ in Eq. (2) defines the size of each constitutive $\mathrm{LG}_{p,l}$ mode, and subsequently, the entire beam. Figure S1 (b-f) shows the amplitude and phase distributions for different values of $w$ with the beam waist $w_0 = w$. They are identical to ones obtained from Eq. (S7).

## S3. Trefoil Knot Coefficients Optimization

The value for the parameter $w$ that performs optimally in the presence of turbulence must be chosen, as this parameter dictates how isolated the knotted structure is from the outer singularities [1]. The beam waist $w_0$ was kept constant in this study. For SR = 0.95 we found that $w = 1.125$ leads to the best stability of the knot in turbulence (Fig. S2). When $w < 1.125$ the outer singularities move closer to the isolated knot and the probability of the singularity lines reconnection in turbulence increases. Even though setting $w > 1.125$ results in a completely isolated knot, the overall knotted structure becomes small, thereby increasing the probability of a reconnection event occurring inside the knot.

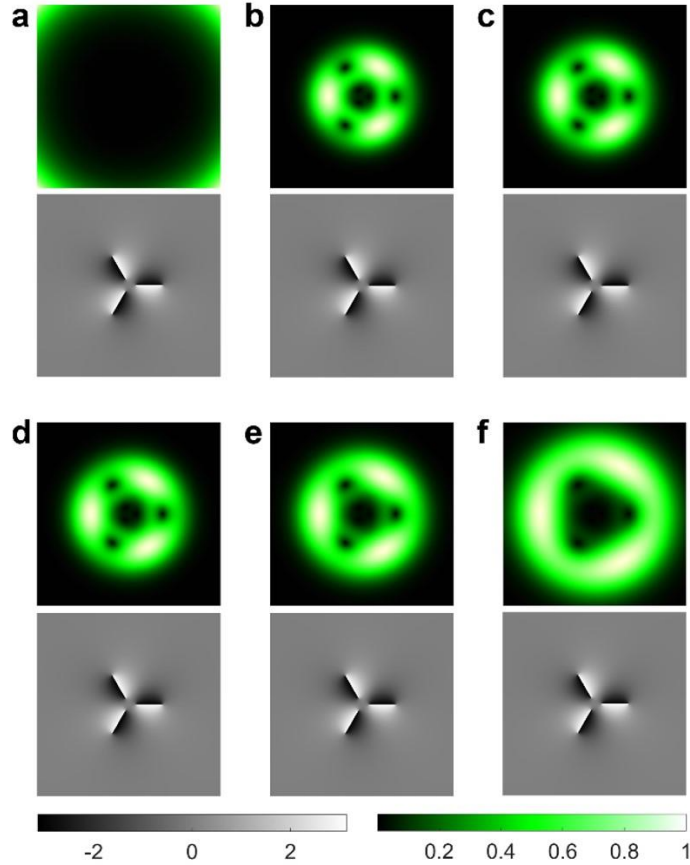

**Figure S1**. Amplitude and phase distributions for (a) the Milnor polynomial expression in Eq. S1 and (b)-(f) after including the Gaussian envelope with different widths $w$ at $z = 0$ (b) $w = 1$, (c) $w = 1.05$, (d) $w = 1.1$, (e) $w = 1.2$, and (f) $w = 1.5$.

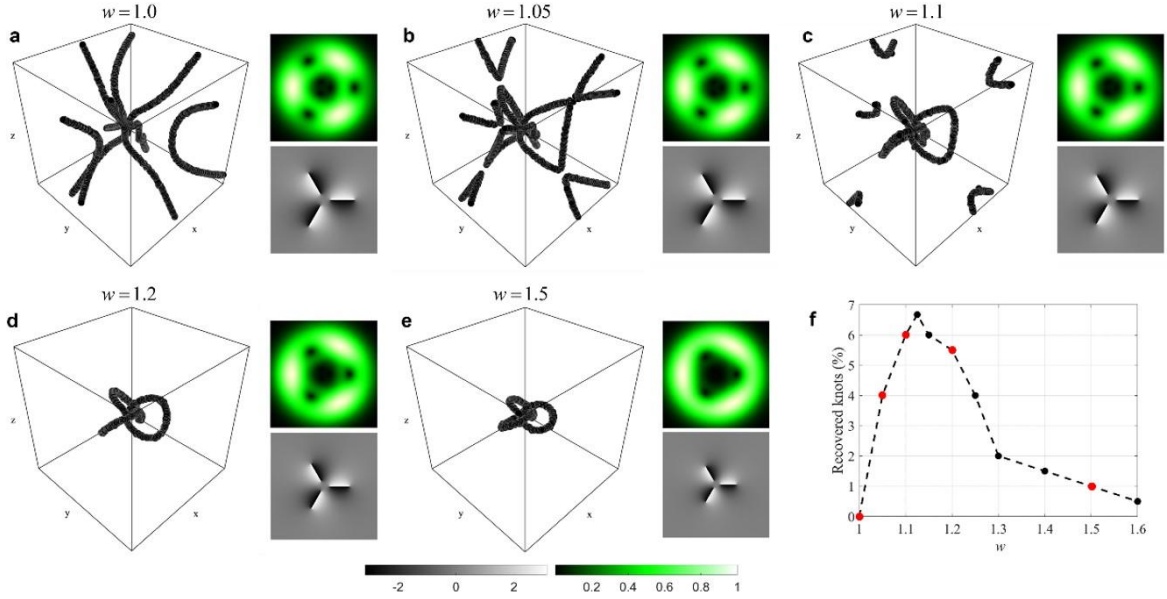

**Figure S2**. Examples of a trefoil knot using Eq. S3 with (a) $w = 1$, (b) $w = 1.05$, (c) $w = 1.1$, (d) $w = 1.2$, and (e) $w = 1.5$ with their respective intensity and phase distributions. (f) Curve showing the percentage of recovered trefoils for various Gaussian envelope widths at SR = 0.95.

## S4. Measuring optical knots in turbulence

Figure S3 summarizes the steps for the reconstruction of the 3D knot from experimental data. Following the measurement of the complex field (shown in Fig. S3 (a)) in a particular plane $z = 0$ (in the middle of the Rayleigh length $z_R = kw_0^2$), we numerically propagate this field in the positive and negative z-direction using an angular spectrum algorithm [2]. Note that the developed procedure can be used to reconstruct the 3D knot from any plane, *i.e.*, not necessarily from the plane corresponding to $z = 0$. The 3D intensity resulting from this reconstruction procedure is shown in Fig. S3 (b). The singularities in the *x-y* plane are found by determining the points corresponding to $\text{Re}\{E\} = \text{Im}\{E\} = 0$, where $E$ is the complex field at every longitudinal position. Figure S3 (c) shows the reconstructed knotted structure. To further improve the precision of finding the locations of the singularities, we repeat the same procedure and search for the singularities in the *x-z* and *y-z* cross-sections, as shown in Fig. S3 (d). Because the mesh is finite, the position of singularities obtained from different planes may yield slightly different coordinates. This results in multiple, closely spaced points of singularities along the same singularity line (as shown in Fig. S3 (e)). Therefore, we are discarding some of these points to reduce the density of dots, thereby accelerating subsequent steps. The resulting knot is shown in Fig. S3 (f). Now, we establish the correct trajectory of the retrieved singularities using the Traveling Salesman problem, leading to the structure in Fig. S3 (g). Finally, in Fig. S3 (h), interpolation is applied using a polynomial approximation to smooth out the overall singularity lines.

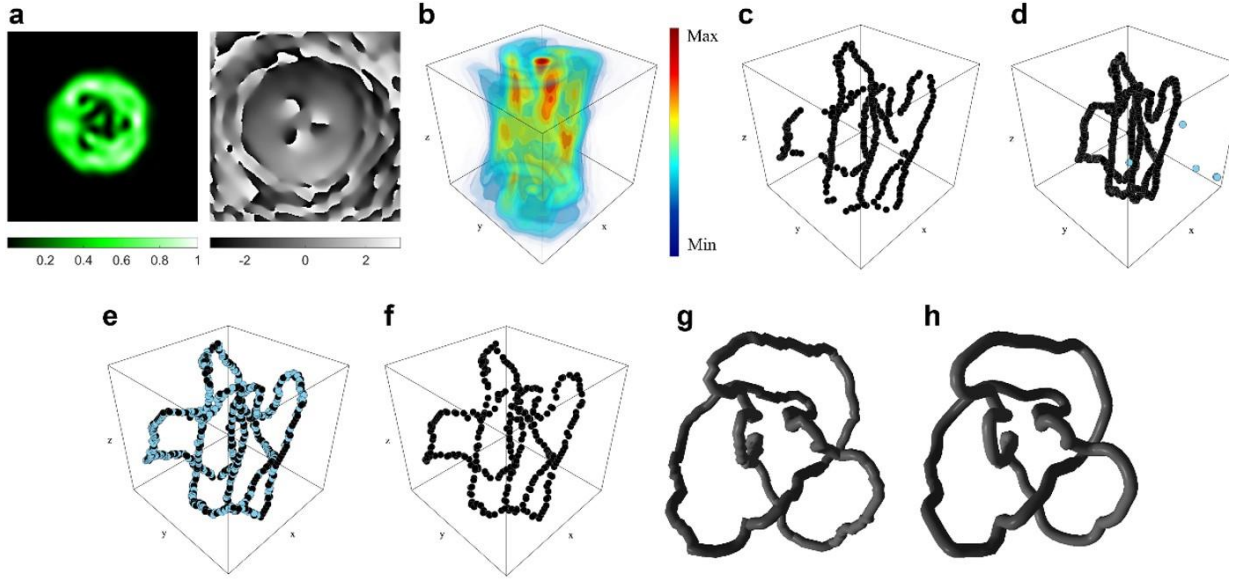

**Figure S3**. (a) Experimentally measured intensity and phase distribution of a trefoil knot after interaction of turbulence with $SR = 0.95$, (b) Intensity distribution after post- and back-propagating the measured optical field. (c) Phase singularities at *the x-y* cross-section at every longitudinal position. (d) Phase singularities after scanning the *x-z* and *y-z* cross-sections. (e) Recovered singularities with chosen dots highlighted (blue) to be discarded in order to increase the efficiency of the algorithm. (f) Resulting knotted structure after removing the dots highlighted in panel (e). (g) trefoil knot after using a Traveling Salesman problem to predict the correct singularity trajectory. (h) Final structure after interpolation to smooth the measured phase singularity trajectories.

## S5. Hopf link

Here, to demonstrate the generality of our approach, we also apply the same procedure to generate other types of knots. We present the results concerning Hopf links. After projecting the Milnor polynomial onto the LG basis, the coefficients for the Hopf link are derived as follows:

$$\overbrace{1 - 2w^2 + 2w^4}^{(p,l)=(0,0)} \bigg| \overbrace{2w^2(1 - 2w^2)}^{(1,0)} \bigg| \overbrace{2w^4}^{(2,0)} \bigg| \overbrace{4\sqrt{2}w^2}^{(0,2)}. \tag{S17}$$

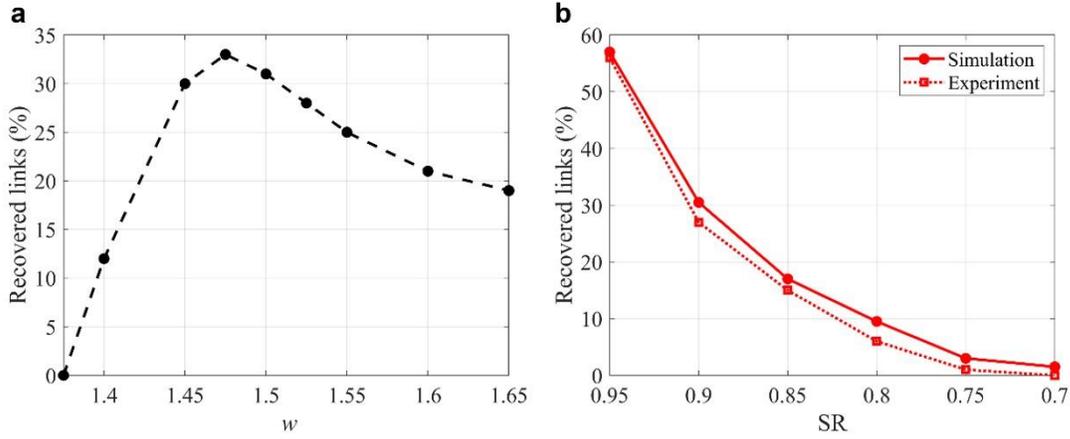

**Figure S4**. (a) Percentage of recovered hopf links using the expression obtained directly from the Milnor polynomial. (b) Percentage of recovered Hopf links from simulation and experiment after optimization.

Figure S4 shows the calibration of the $w$ parameter (a), and the percentage of recovered Hopf links after turbulence (b) for $w = 1.475$. The coefficients for each $(p,l)$ mode used in the experiment are: (0,0) 2.96, (1,0) -6.23, (2,0) 4.75, and (0,2) 5.49.

## S6. More examples of recovered topologies

Figure S5 presents additional examples of measured trefoil knots, Hopf links, and unknots retrieved after interaction with turbulence (SR = 0.85). Furthermore, Fig. S6 demonstrates extreme cases where the three-dimensional knotted structure is completely destroyed. It is important to note that these obliterated topological structures are a combination of one or more unknots and/or extra singularity lines, which reconnect to those associated with the Gaussian envelope extending from $-\infty$ to $\infty$ (refer to Section S3 and Fig. S2). This reconnection occurs due to turbulence, leading to the loss of the original topological information. In this work, we have classified these destroyed structures as unknots.

## S7. Limitations of the method

The concept of the optimization process was inspired by Ref. [1], where the authors demonstrated that a numerical adjustment of the coefficients can enhance the precision of the measurements. It is noteworthy that the percentage of recovered Hopf links is lower than the percentage of recovered trefoil knots, which is somewhat counter-intuitive. Typically, higher-order topological structures tend to be less stable under the presence of external perturbations. For instance, crosstalk increases for higher-order vortices, making them less robust under turbulence. This counter-intuitive result can be understood as follows. Comparing Eqs. (S9) and (S17), we notice that higher order knots are built of a superposition of a larger number of $LG_{p,l}$ modes. The larger number of constitutive modes leads to more degrees of freedom for the numerical optimization, and therefore, to more stable results. The results presented in this study could potentially be further improved by adding more radial modes to the knotted field.

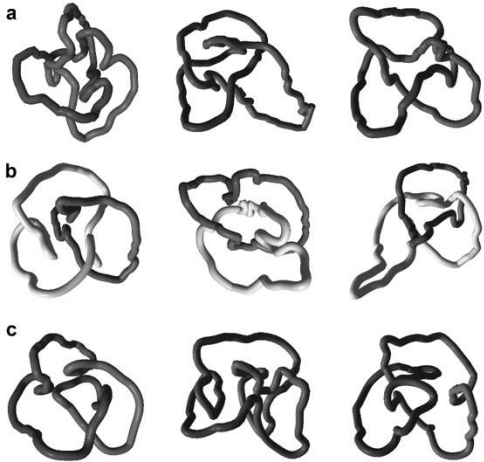

**Figure S5**. (a) Trefoil, (b) Hopf, and (c) unknot structures recovered from the experiment for SR = 0.85.

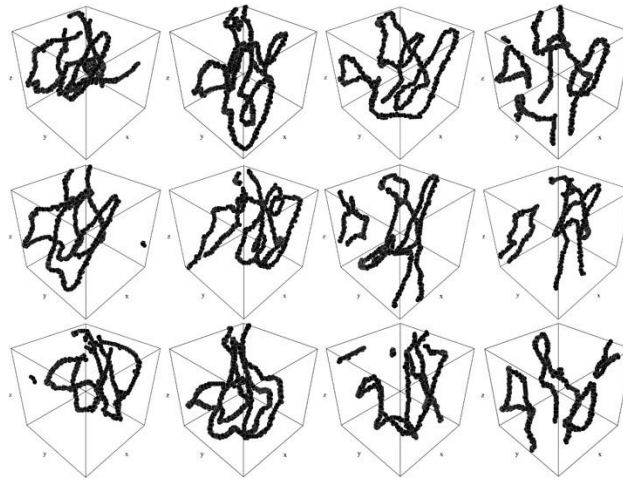

**Figure S6**. Examples of destroyed structures recovered from the experiment with SR = 0.9.

## Supplementary References

1. Dennis, M.R., et al., *Isolated optical vortex knots.* Nature Physics, 2010. **6**(2): p. 118-121.
2. Zhong, J.Z., et al., *Reconstructing the topology of optical vortex lines with single-shot measurement.* Applied Physics Letters, 2021. **119**(16).